\documentclass[twocolumn,aps,prl,superscriptaddress]{revtex4}

\usepackage{graphicx,color,hyperref}

\usepackage{amsmath}
\usepackage{amsfonts}
\usepackage{amssymb}

\usepackage{braket}

\newcommand{\tr}{\mathrm{tr}}

\usepackage[normalem]{ulem} 

\begin{document}

\title{
Maximum entropy quantum state distributions
}

\author{Alexander Altland}
\affiliation{Institut f\"ur Theoretische Physik, Universit\"at zu K\"oln, Z\"ulpicher Str. 77, 50937 Cologne, Germany
}

\author{David A. Huse}
\affiliation{Department of Physics, Princeton University, Princeton, NJ 08544, USA}
\affiliation{Institute for Advanced Study, Princeton, NJ 08540, USA}

\author{ 
Tobias~Micklitz 
}
\affiliation{Centro Brasileiro de Pesquisas F\'isicas, Rua Xavier Sigaud 150, 22290-180, Rio de Janeiro, Brazil 
} 

\date{\today}

\begin{abstract}
We propose an approach to the realization of many-body quantum state distributions  inspired by combined principles of thermodynamics and mesoscopic physics. Its essence is a maximum entropy principle conditioned by conservation laws.  We go beyond traditional thermodynamics and condition on the full distribution of the conserved quantities. 
The result are quantum state distributions whose deviations from `thermal states'  get  more pronounced in the limit of wide input distributions. We describe their properties in terms of  entanglement measures and discuss strategies for state engineering by methods of current date experimentation. 
\end{abstract}

\maketitle

In the early nineteenth century, thermodynamics was introduced to  describe the
state of complex systems on the basis of a minimal specification. The provided
information would specify the values of certain macroscopic observables, such as the
energy density. In the language
of statistical thermodynamics, the \emph{minimal} amount of information
provided translates to \emph{maximal} entropy of the remaining unspecified microscopic properties. 
For example, the canonical distribution
$\rho=Z^{-1}\exp(-\beta H)$ defines a distribution for a system's micro-states
maximizing entropy under the condition of fixed average energy. 
The entropy, conditioned by the specified information, is $S=-\tr(\rho \ln \rho)$.

Two centennials later, the principles of that approach have not lost their power.
However, progress in experimentation now makes it possible to apply them to quantum systems of \emph{mesoscopic} proportions: systems large enough to be efficient information scramblers and `thermalizing', yet small enough to harbor large quantum fluctuations. Under these conditions, it becomes natural to generalize the specified information from given average values of conserved quantities to full statistical distributions. For example, for  a spin chain with
conserved total $z$-axis magnetization, $S_z$, one may consider initial pure states $\hat \rho_0=|\Psi_0\rangle\langle \Psi_0|$ engineered to realize a chosen magnetization distribution $p(M)\equiv  \langle  \Psi_0 |
\delta_{\hat S_z,M}
|\Psi_0\rangle$. Under dynamical evolution by a Hamilton operator $\hat H(t)$ --- assumed to be many-body chaotic --- $\Psi_0$ 
evolves into a pseudo-random pure state, $\hat \rho=|\Psi\rangle\langle\Psi|$, with the same  $p(M)$ (since $[\hat{H}(t),S_z]=0$), and perhaps other constraints, such as a conserved total energy. (By contrast, basic  thermodynamics would be
content with fixing just $\langle M\rangle=\langle \Psi|\hat S_z|\Psi \rangle$.) 

In this letter we fuse concepts of thermodynamics and quantum mesosocopics. We
consider pure states defined to maximize entropy, conditional that they realize
a  distribution $p(Q)$ of a scalar `charge' , $Q$, --- energy, particle number, uni-axial magnetization, etc.  
We describe the physical properties of such
states, in particular their entanglement properties.  The limiting case of $P(Q)=\delta_{Q,Q_0}$ models eigenstates of $\hat{Q}$ 
satisfying an eigenstate thermalization hypothesis (ETH)~\cite{Deutsch1991,Srednicki1994,Rigol2008}. 

The key input determining the entanglement properties of our states are deviations between the specified distribution $p(Q)$ and the spectral distribution, $\Omega(Q)$, of the
conserved operator, $\hat Q$, where $D\Omega(Q)$ is the number of $Q$-eigenvalue states in a $D$-dimensional Hilbert space. 
The unconstrained case, $p(Q)=\Omega(Q)$, has the highest entanglement, corresponding to Page's maximally random pure states~\cite{Page93}.  Constraining to eigenstates of $\hat{Q}$ reduces the entanglement, but much larger reductions can be obtained by specifying 
distributions broad compared to $\Omega(Q)$. Within a parameter space spanned by the center and width of $p(Q)$, we will discuss various reference configurations, among them  Page's random states~\cite{Page93} (no input information
provided),  thermal distributions~\cite{Sugiura12,Sugiura13,Sugiura18}, microcanonical
distributions~\cite{VidmarRigol17,Rigol21}, and very broadly distributed $p(Q)$. We will argue that for spin systems of mesoscopic proportions the engineering of the latter is in experimental reach and should thus lead to tunable non-thermal signatures in, e.g., 
R\'enyi entropies~\cite{islam2015measuring,Kaufman2016}
or spin-correlation probes.

{\it Maximum entropy distributions:---} Without much loss of generality, we consider a $N$-qubit system, and in it a conserved operator which is subsystem additive: a partition of the system into two
subsystems $A$ and $B$ of size $N_A$ and $N_B=N-N_A$ implies a decomposition $\hat{Q}=\hat{Q}_A+\hat{Q}_B$. On the same basis, $\hat
Q$'s eigenstates, $|n\rangle$ are labeled by a $D=2^{N}=2^{N_A+N_B}=D_AD_B$ dimensional index $n=(a,b)\in \mathbb{Z}_2^N$ with eigenvalues $Q_n$. 

 Assuming a
given charge distribution, $p(Q)$, we now construct a distribution
$P(\Psi)=P(\{\Psi_n\})$ defined on the space of pure states satisfying two
conditions: first, in the limit of large $N$, states drawn from $P$ satisfy the condition $p(Q)= \langle \Psi|\delta_{\hat
Q,Q}|\Psi\rangle$, where $\delta_{\hat Q,Q}$ is a $\delta$-function or a Kronecker
$\delta$, depending on whether $Q$ is continuous or discrete. Second, the information entropy $S[P]=-\langle \ln(P)\rangle_P $ be maximal, where we introduced the shorthand notation $\langle \ldots\rangle_P=\int D\Psi P(\Psi)(\ldots)$. A distribution satisfying these criteria is found by extremizing the functional  
  \begin{align}
    \label{eq:ActionFunctional}
     A[P]
     \equiv 
     S[P]
     &+
     \lambda \left(\langle 1\rangle_P -1\right)
     +
     \lambda_0 \left( 
     \langle\langle \Psi| \Psi\rangle\rangle_P - 1\right)
     \nonumber\\
     & +
     \sum_Q \lambda(Q) \left( \langle     \langle \Psi| \delta_{\hat Q,Q} |\Psi\rangle\rangle_P - p(Q) \right).
 \end{align}
Here, the Lagrange multipliers, $\lambda_0 $ and $\lambda(Q)$ impose state normalization, $\langle\langle \Psi| \Psi\rangle\rangle_P = 1$, and  the distribution property,   $\langle     \langle \Psi| \delta_{\hat Q,Q} |\Psi\rangle\rangle_P = p(Q)$,  only on average over the distribution $P[\Psi]$. In exchange for the typicality assumption $\langle F[\Psi]\rangle_\Psi\approx F[\Psi]$, which we assume holds true in the limit of large Hilbert space dimension, we obtain an extremization problem that is easy to solve:

In the supplemental material~\cite{supplemental_material}, we show that the straightforward variation of the action functional yields the solution
\begin{align}
   \label{eq:Distribution}
    &\frac{D}{\lambda_0+\lambda(Q)}=\frac{p(Q)}{\Omega(Q)},\qquad \sum_Q \frac{D\Omega(Q)}{\lambda_0+\lambda(Q)}=1,\cr
    &\rho_n\equiv \langle |\Psi_n|^2\rangle= \frac{p(Q_n)}{D \Omega(Q_n)}.
 \end{align} 
The first two of these equations fix the Lagrange multipliers, and the third states that the distribution of wave function amplitudes is Gaussian, with a variance set by the specified charge distribution $p(Q)$.  With  $\rho_n\ge 0$ and $\sum\rho_n=1$, we interpret $\rho =\{ \rho_n\}$ as the  average distribution of spectral weight over Fock space (see table \ref{table:1} for an overview of the various distributions relevant to our discussion.)

To gain some familiarity with these expressions, consider the limiting case of unconstrained states, $\lambda(Q)=0$. The above equations are then solved by $\lambda_0=D$. For this value, $\Psi_n$ are Gaussian variables with uniform variance $D^{-1}$; the  random state vectors considered by Page. In this particular case, the charge distribution $p(Q)=\Omega(Q)$ is dictated by the native spectral distribution. In the following, we consider what happens if we condition to Gaussian charge distributions \begin{align}
  p(Q)
 = \frac{1}{\sqrt{2\pi}\Delta Q} \exp\left(-\frac{(Q-\bar{Q})^2}{2(\Delta Q)^2} \right),
\end{align}
of general center $\bar{Q}$ and width $\Delta Q$. 
The observation that the limiting case $p(Q)=\Omega(Q)$ corresponds  to uniformly distributed states of minimal structure suggests to quantify the `input information' provided by $p(Q)$ in terms of the statistical distance to $\Omega(Q)$, i.e. the Kullback-Leibler divergence $I_\mathrm{in}\equiv -D_\mathrm{KL}(p||\Omega)=\sum_Q p(Q)\ln \left(\frac{p(Q)}{\Omega(Q)}\right)$.  Using
 Eqs.\eqref{eq:Distribution}, it is
 straightforward to verify that
 \begin{align}
     I_\mathrm{in}=-S(\rho)+\ln D,
 \end{align}
where $S(\rho)=-\sum_n \rho_n \ln(\rho_n)$ is the von Neumann entropy of the distribution $\rho$. 
In the following, we calculate the effects of this specified charge distribution $p(Q)$ on 
the entanglement entropies defined by subsystem partitions.

\begin{table}[t!]
\begin{center}
\begin{tabular}{c|c|l}\hline
\textbf{input} & $\Omega(Q)$ & spectral distribution of charge\cr 
& $p(Q)$ & engineered distribution of charge\cr \hline 
\textbf{output} & $P(\Psi)$& distribution of pure states conditioned via $p$\cr 
& $\rho_n$ & spectral weight distribution in Fock space\cr 
& $\rho_{A,a}$ & subsystem reduction of $\rho$\cr 
& $p_A(Q_A)$ & charge distribution over subsystem \cr 
\hline 
\end{tabular}
\end{center}
\caption{Overview over the different statistical distributions defined by a conserved operator $\hat Q$ in a many body system: the eigenvalues of $\hat Q$ are distributed according to $\Omega(Q)$; we are interested in states redistributing charge into $p(Q)$. Embedding this condition with the maximum entropy principle, we obtain average pure state distributions $\rho_n$, and their subsystem marginals $\rho_{A,a}$ for $n=(a,b)$. Finally, $p_A(Q_A)$ describes the average distribution of charge on the subsystem $A$.}
\label{table:1}
\end{table}

{\it Entanglement entropies:---} We cut our system into two pieces, $A$, $B$, with
associated Fock spaces $\mathcal{F}=\mathcal{F}_A\otimes \mathcal{F}_B$ of  dimensions $D_A\le D_B$. For a given pure state $\hat \rho=|\Psi\rangle
\langle \Psi|$, we consider the  reduced density matrices $\hat \rho_A=\tr_B (\hat
\rho)$, and entanglement entropies $S=-\langle \mathrm{tr}(\hat \rho_A \ln \hat \rho_A)\rangle_\Psi \simeq - \mathrm{tr}( \rho_A \ln  \rho_A)\equiv S^\mathrm{a}$, where in the `average' entropy, $S^\mathrm{a}$, we assume self averaging $\hat \rho_A\simeq \langle \hat \rho_A\rangle_\Psi\equiv \rho_A$. Page has shown that in the absence of conditioning $S^\mathrm{a}=S_\mathrm{th}\equiv N_A \ln 2$ is trivially thermal. In the following we show how this changes due to specifying $p(Q)$. 
For a basis decomposition $n=(a,b)$, $\rho_A=\{ \rho_{A,a}\}$ is described by a list of $D_A$ coefficients $\rho_{A,a}$.
Using the definition of $S_A^\mathrm{a}$ in combination with Eq.~\eqref{eq:Distribution}, we obtain 
\begin{align}
  \label{eq:AverageEntropy}
&\Delta S^\mathrm{a}
=
-\sum_{Q_A}p_A\ln\left(\frac{p_A}{\Omega_A}\right) =-D_\mathrm{KL}(p_A||\Omega_A), \cr 
&\qquad p_A(Q_A)
=
\Omega_A(Q_A)\sum_{Q_B}\frac{\Omega_B(Q_B) p(Q_A+Q_B)}{\Omega(Q_A+Q_B)},
\end{align}
for the difference $\Delta S^\mathrm{a}=S_A^{\rm av}-S_\mathrm{th}$. Conceptually, $p_A(Q_A)$ describes the unit normalized~\cite{footnote2} 
charge distribution imprinted on subsystem $A$ via the input charge distribution $p(Q)$ in the maximum entropy ensemble.  Eq.~
\eqref{eq:AverageEntropy} states that the resulting reduction of the entanglement entropy of the average density matrix equals the 
(Kullback-Leibler) deviation of the induced distribution from the spectral distribution of the operator. 

To make the  result Eq.~\eqref{eq:AverageEntropy}  concrete, we need to evaluate
the convolutions in Eq.~\eqref{eq:AverageEntropy} in more explicit terms. Reflecting the local additivity principle obeyed by the variable $Q$, we assume that, except in the far tails of the spectrum, the density of states is Gaussian
\begin{align}
\label{eq:GaussianDoS}
  \Omega(Q)=\Omega(Q,N)=\frac{1}{\sqrt{2\pi}\Gamma}e^{-\frac{Q^2}{2\Gamma^2}}\equiv \frac{1}{\sqrt{2\pi N}\gamma}  e^{-N\frac{q^2}{2\gamma^2}},
\end{align}
where we define the $Q$-variable such that $Q=0$ corresponds to the center of the distribution $\Omega(Q)$, and the scaled variables $Q=Nq$, $\Gamma=\sqrt{N}\gamma$. The same function describes the subsystem spectral densities as $\Omega_X(Q_X)=\Omega(Q_X,N_X)$, $X=A,B$.

\begin{figure}[t!]
\centering
\includegraphics[width=8.5cm]{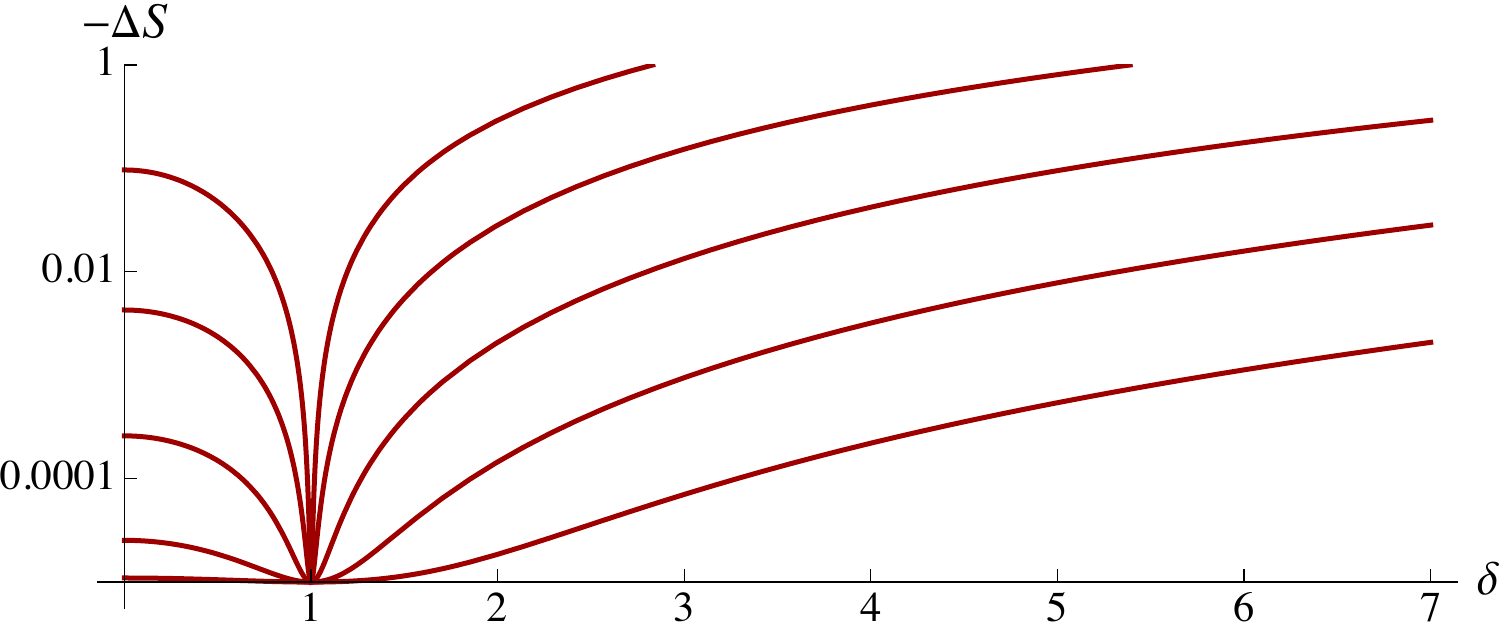}
\caption{\label{fig:EntanglementEntropy}
Entropy Eq.~\eqref{eq:AverageEntropy} plotted as a function of $\delta$ for $n_A=2^{-9},2^{-7},\ldots 2^{-1}$, with larger entropies corresponding to larger subsystem sizes.  
}
\end{figure}

In the supplemental material~\cite{supplemental_material} we show that for these spectral densities the reduced charge density assumes the form
\begin{align}
\label{eq:ReducedChargeDensity}
   p_A(Q_A)&\approx   \frac{N \sqrt{N_B}}{\Gamma\sqrt{2\pi N_A }}\int dx\, e^{-\frac{N_AN_B}{2\Gamma^2}x^2}p(q_A N+xN_B), 
 \end{align} 
i.e. the input charge distribution convoluted against a smoothing function which knows about the spectral distribution of $\hat Q$ and the relative sizes of the subsystems. In the limit of large system sizes and fixed $N_A/N_B$, the variable $x\sim
N^{-1/2}$, and so $x N_B\sim N^{1/2}$, which is small compared to the variable
$q_A N\sim N$. For a broad 
input distribution, $p$, the $x$-dependence of $p$ becomes negligible, and the $x$-integral yields $p_A(Q_A)\approx  \frac{N}{N_A}p(Q_A (N/N_A))$.
This result states that the charge distribution on $A$ inherits that of the system at
large. In the opposite limit of a sharply-peaked input distribution, 
$p(Q)=\delta(Q-\bar Q)$,  Eq.~\eqref{eq:ReducedChargeDensity} 
collapses to $p_A(Q_A)\sim \exp(-\frac{N_A}{2\Gamma^2 N_B}(q_AN -\bar Q)^2) $. This broadening of the $\delta$-function 
occurs due to the subsystem trace, which reduces the degree of constraint. 
More generally, for a Gaussian distribution of width $\Delta Q$ the 
computation of the entanglement entropy Eq.~\eqref{eq:AverageEntropy} (see 
supplemental material~\cite{supplemental_material}) yields 
\begin{align}
  \label{eq:OutputInformation}
  \Delta  S^{\mathrm{a}}&=-\frac{n_A}{2}\left(\delta^2+\kappa^2-1\right) +
 \frac{1}{2}\ln\left( 1+n_A \left(\delta^2-1\right)\right),
 \nonumber\\ 
 &n_A=\frac{N_A}{N}, \qquad \delta=\frac{\Delta Q }{\Gamma},\qquad \kappa=\frac{|\bar Q|}{\Gamma}. 
\end{align}
According to this result, the state information output via the average entanglement
entropy depends on the three parameters: relative system size, $n_A$; and deviation,
$\kappa$, and width, $\delta$, of the input distribution relative to the width of
$\hat Q$'s spectral density. Figure \ref{fig:EntanglementEntropy} shows this quantity
for a centered distribution, $\kappa=0$, as a function of the parameter $\delta$ for multiple values of $n_A$. Its most striking feature is the non-monotonic
dependence, with a minimum $\Delta S^\mathrm{a}=0$ for $\delta=1$, or $\Delta Q=\Gamma\propto
N^{1/2}$ equal to the width of the native spectral distribution $\Omega(Q)$. 
The entanglement entropy may be reduced by sharpening the input distribution (thus increasing
the specified information), with a limit $\Delta S^\mathrm{a}= (n_A+\ln{(1-n_A)})/2$ for $\delta \rightarrow 0$. However, perhaps
unexpectedly, 
output information also results from  \emph{broadening} the input. 
Already for distributions with $\delta \approx \sqrt{2}$, widening becomes a stronger information booster than sharpening.  For extensively wide distributions, $\delta \sim \sqrt{N}$, or $\Delta Q \sim N$, the anomalous contribution to the entanglement entropy is likewise extensive, $\Delta S^\mathrm{a}\approx - n_A \delta^2\sim N_A$. (The extreme limit within this class of distributions is realized for a flat input $p(Q)=\mathrm{const}$. In this case, a straightforward estimate yields $\Delta S^\mathrm{a}\approx -N_A/24$, respecting the positive definiteness of the total entropy $S=S_\mathrm{th}+\Delta S^\mathrm{a}$.)

\noindent {\it Fluctuation corrections:---} The above construction describes the entropy of the averaged state over a wide range
of parameters. Quantitative corrections  to Eq.~\eqref{eq:AverageEntropy} arise for
average values $\bar Q$ in the far tails of the distribution $\Omega(Q)$, where the
Gaussian distribution breaks down, or for input densities $p(Q)$ which cannot be
modeled as locally Gaussian around a single maximum. While we do not consider such
effects in the present paper, there remains one open conceptual point namely that the
average entanglement entropy of pure states, $S\not=S^\mathrm{av}$ differs from that
of the state-average. For example, Page has shown~\cite{Page93} that in the absence of
conditioning, $\Delta S^\mathrm{a}=0$, while for the full  entanglement entropy,  $\Delta
S= - \frac{D_A}{2D_B}$ is exponentially small for asymmetric cuts,
$D_A/D_B=2^{N_A-N_B}$, but becomes sizeable for near-equal subsystem sizes (see also ~\cite{Foong1994,SanchezRuiz1995}).

\begin{figure}[t!]
\centering
\includegraphics[width=8.5cm]{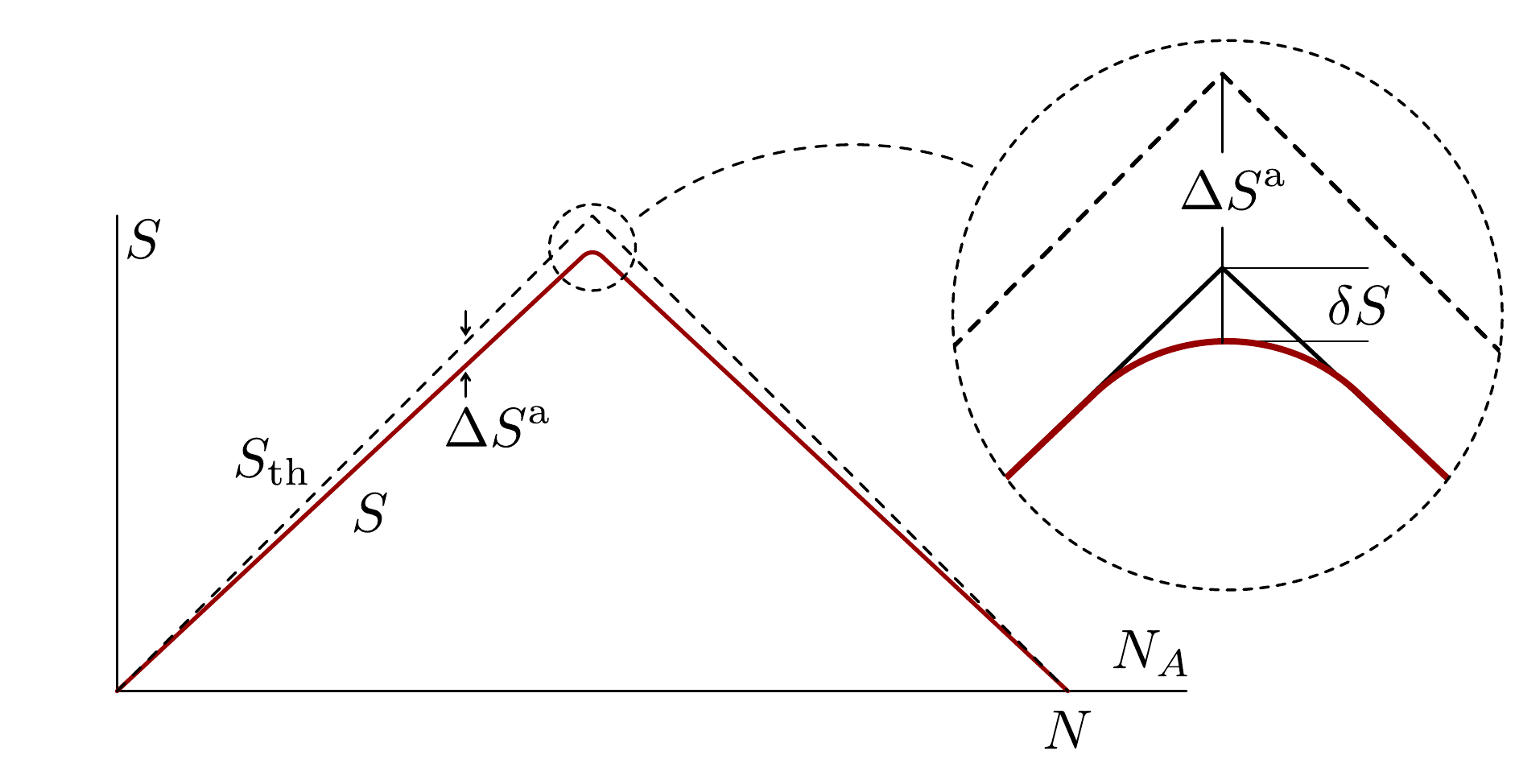}
\caption{\label{fig:EntropyWedge}
Schematic structure of the entanglement entropy as a function of subsystem size, $N_A$. Discussion, see text.  
}
\end{figure}

In order to compute (quantum) fluctuation corrections to the result
 Eq.~\eqref{eq:AverageEntropy}, we need to go back to a first principles
 representation $S=-\partial_r|_{r=1}M^r$ of the entanglement entropy in terms of the
moments $M^r=\langle \mathrm{tr}_A(\hat \rho_A^r)\rangle_\Psi$ of the reduced density matrix, 
$\rho_{A,aa'}=\sum_b  \Psi_{(ab)}\bar\Psi_{(a'b)}$~\cite{Stanford2019,Liu20}. For the Gaussian
distribution Eq.~\eqref{eq:state_distribution}  this expression becomes a sum over all combinatorial pairings of wave function amplitudes, where so far we considered the nearest neighbor pairings, $\langle
 \Psi_{(ab)}\bar\Psi_{(a'b)}\rangle_\Psi$. In 
 the supplementary material~\cite{supplemental_material} we consider
 the contribution of other pairings for the exemplary case of the microcanonical
 distribution $p(Q)=\delta_{Q,\bar Q}$. (The computation for generic distributions is
 more complicated but does not lead to qualitatively different conclusions although details change.) 
 It turns
 out that  for $N_A\ll N_B$ and charges $\bar{Q}$ away from the extreme tails, the average entropy $S^\mathrm{a}$ approximates the full
 one up to corrections in  $D_A/D_B$. (The scaling in Hilbert space dimensions follows
 from the observations that for Gaussian contractions different from the average one
 one needs to pay in free summations over the $B$-subspace.) For $N_A\approx N_B$, we
 obtain a correction, $\delta S$, required to establish the symmetry of the
 `entanglement wedge' (see Fig.~\ref{fig:EntropyWedge}) under exchange
 $A\leftrightarrow B$. 
 Technically, the wedge 
 reflects zero singular values in the pure states Schmidt decomposition.
  For Page states these zeros only occur for $N_B>N_A$, however, it is more subtle in systems with conserved charges. 
 In the supplemental material~\cite{supplemental_material} we show that for 
 microcanonical distributions 
  the `wedge' is smoothened 
 by a correction $\delta S \propto -\sqrt{N}\bar{q}$ (as first noted in~\cite{VidmarRigol17}). Higher order corrections
 in $D_A/D_B$ also considered by Page, on the other hand, remain bounded by $-1/2$ (see  supplemental material~\cite{supplemental_material} 
  for explicit expression). 
The upshot of this
 discussion is that for asymmetric cuts the average entropy approximates the full
 one, and even for equipartition
 $N_A=N_B$ represents a good approximation.

\noindent \emph{State preparation: ---} The above discussion has shown how maximum
entropy states realizing broad charge distributions differ from thermal states. For
systems in the thermodynamic limit $N\to \infty$, charge states lie with overwhelming
probability inside the $\sim \sqrt{N}$ tolerance window of the native distribution
$\propto D\Omega(Q)$, and the engineering of pure states of broader distribution may
be hard. However, for the mesoscopic values of $N$ realized in current date AMO 
experiments or numerical analysis, broad distributions may be (perhaps even
inadvertently) realized. To give an example, we consider a system of $N$ spin $1/2$
with assumed conservation of the total $z$-axis spin component,  $\hat Q =\hat S_z$.
Assume the system divided into $M=N/L$ blocks of $L$-sized cat states~\cite{johnson2017ultrafast,Gao10qubit2010,yao8qubit2012,Wang2016,Hacker2019,Duan2019,Omran2019}
$|\Psi_a\rangle=\frac{1}{\sqrt{2}}(|\uparrow\rangle_L + |\downarrow\rangle_L)$
(with $|\uparrow,\downarrow\rangle_L$ the $\hat Q$-eigenstates at values $S_z=\pm L/2$), and 
consider $|\Psi\rangle\equiv \otimes_{a=1}^M |\Psi_a\rangle$. For this state, a
straightforward calculation shows that $p(Q)=\langle \Psi|\delta_{Q,\hat
S_z}|\Psi\rangle=\frac{1}{2^M}\sum_{a=0}^M\binom{M}{a}\delta_{Q,(a-\frac{M}{2})L}=\frac{1}{2^M}\binom{M}{\frac{M}{2}+\frac{Q}{L}}\simeq
\frac{\sqrt{2}}{\sqrt{\pi M}}\exp(\frac{2Q^2}{ M L^2})$. This particular state has charge distributed over a range $\sim L \sqrt{M}=N/\sqrt{M}$. This means that for fixed block number, $M\sim N^0$, we have an extensively broad distribution. In this scaling regime, the above product cat will evolve under chaotic time evolution into  random states different from structureless ETH states. These differences show in the entanglement properties emphasized in this paper, but more directly also in magnetization probes revealing the dynamical conservation of the charge distribution despite the systems many body chaotic dynamics.

{\it Discussion:---} We introduced an approach to quantum state design which combines
a maximum entropy principle with premeditated distributions of dynamically conserved
observables in meso-sized many body quantum systems. This distribution survives
unmasked by the chaotic dynamics and shows at various levels of the system
description, beginning with the interpretation of the states themselves: for a given
$p(Q)$, we defined a distribution of states $P(\Psi)$, such that individual states
$\hat \rho =|\Psi\rangle \langle \Psi|$ drawn from it, generate the input in the
sense $p(Q)=\mathrm{tr}(\hat \rho \delta_{\hat Q,Q})$. (As an alternative to this
pure state interpretation, one may also consider the mixed state $\hat
\rho'=\sum_\Psi P(\Psi)|\Psi \rangle\langle \Psi|$, which has the same property
$p(Q)=\mathrm{tr}(\hat \rho \delta_{\hat Q,Q})$, see   supplemental material~\cite{supplemental_material} 
for more on the comparison of the two views.) When a component $B$ of the system is
traced out and in this way becomes part of an environment,  an induced charge
distribution $\rho_A(Q_A)$  emerges from the given one by a convolution over a kernel
obtained from $\hat Q$'s spectral distribution, Eq.~\eqref{eq:ReducedChargeDensity}.
This effective distribution is broader than the original one, reflecting the loss of
information due to the partial trace. The statistical distance of the induced
distribution to the spectral density of the conserved operator in $A$ determines the
entanglement entropy of our quantum states, Eq.\eqref{eq:AverageEntropy}. This
central result of our paper quantifies the difference to thermal states, a principal
observation being that broad input distributions lead to the strongest deviations
from ETH states. The straightforward generalization of this finding to multiple but mutually commuting conserved quantities (such as energy and particle number) shows that in this case the individually computed entropies $\Delta S^\mathrm{a}$ add. 
One point left out here concerns the generalization to distributions of strong non-Gaussianity. (For recent work discussing the influence of Lorentzian distributions on the statistics of quantum states we refer to Ref.~\cite{BogomolgnySieber2018,MonteiroPRL2021})

Our analysis illustrates how entire distributions (and not just single average values) may survive ergodic chaotic time evolution. It would be interesting to see if this principle of state design may be turned into a creative resource. Conversely, one needs to watch out, especially in small or medium sized systems, and check if the application of the ETH hypothesis to a many body state might be incompatible with an (perhaps inadvertently) introduced defined initial charge distribution. 

\emph{Acknowledgments: ---} D.~A.~H. thanks Shayan Majidy and Nicole Yunger Halpern for discussions.  D.~A.~H. is supported in part by NSF QLCI grant OMA-2120757. T.~M.~acknowledge financial support by Brazilian agencies CNPq and FAPERJ. A.~A. acknowledges partial support from the Deutsche
 Forschungsgemeinschaft (DFG) within the CRC network TR 183 (project grant 277101999)
 as part of projects A03. 


\newpage


\section{Supplementary Material}

\subsection{Solution of the variational equations for the state distribution}

We here show how Eqs.~\eqref{eq:Distribution} follow from variation of the functional Eq.~\eqref{eq:ActionFunctional}. The straightforward solution of the variational equation $\delta_{P}A[P]=0$ yields 
$P(\Psi) 
 =
 \frac{1}{Z} e^{-\lambda_0 \langle \Psi|\Psi\rangle 
-\sum_Q \lambda(Q) \langle \Psi|\delta_{\hat Q,Q}|\Psi\rangle}$,
 with a normalization factor $Z$. Using the eigenstate property, $ \langle \Psi|\delta_{\hat Q,Q}|\Psi\rangle=\sum_m |\Psi_m|^2 \delta_{Q_m,Q}$, and integrating over all components $\Psi_{m\not=n}$ but one, we find that  individual amplitudes $\Psi_n$ are Gaussian distributed with 
\begin{align}
\label{eq:state_distribution}
 \rho_n\equiv \langle |\Psi_n|^2\rangle= \frac{1}{\lambda_0+\lambda(Q_n)}. 
\end{align}
This result may now  now be used to fix the Lagrange multipliers
$\lambda_0$ and $\lambda(Q)$. Converting sums into integrals as  $\sum_n F(Q_n)=\sum_Q \Omega(Q) F(Q)$,  the normalization condition yields $1=\langle \sum_n |\Psi_n|^2\rangle=\sum_Q \frac{D\Omega(Q)}{\lambda_0+\lambda(Q)}$. Treating the second condition in Eq.~\eqref{eq:ActionFunctional} in the same  manner, we obtain the first line in Eq.~\eqref{eq:Distribution}. 
Substitution of this result into Eq.~\eqref{eq:state_distribution} yields the second line.

\subsection{The subsystem charge distribution, $p_A(Q_A)$} 
\label{sec:derivation_of_subsystem_charge_distribution_}

In order to derive Eq. \eqref{eq:ReducedChargeDensity}, we represent the subsystem charge distribution as an integral over Gaussian spectral densities,  
\begin{align*}
 &p_A(Q_A)=\frac{\mathcal{N}_A\mathcal{N}_B}{\mathcal{N}_{A\cup B}}N_B \int dq_B\, \cr
 &\quad  e^{-\frac{1}{2\gamma^2}\left(N_A q^2_A+N_B q^2_B-\frac{1}{N}(q_A N_A +q_B N_B)^2 \right)} p(q_A N_A + q_B N_B),
\end{align*}
where we introduced scaled, and effectively continuous variables $Q_X=q_XN_X$, $X=A,B$. Considered as a function of $q_{A,B}$, the exponential function under the integral 
possesses  the saddle points, $q_B=q_A$, indicating that the system favors charge equilibration over its subsystems. With $q_B=q_A+q$, the second order expansion of the integral around this saddle assumes the form Eq.\eqref{eq:ReducedChargeDensity}.

An explicit representation, not relying on the above assumption, may be obtained for a Gaussian input distribution $p(Q)=
  \frac{\Delta Q}{\sqrt{2\pi}} \exp(-\frac{(Q-\bar{Q})^2}{2\Delta Q^2} )$. In this case, the  integral over $q$ yields
  \begin{align*}
    p_A(Q_A)&=\frac{N}{N_A}\frac{1}{\sqrt{2\pi}\Lambda}\exp\left(-\frac{(Q_A \frac{N}{N_A} -\bar Q)^2}{2\Lambda^2}\right),\cr 
    &\qquad   \Lambda^2 = \Gamma^2\frac{ N_B}{ N_A}+\Delta Q^2.
  \end{align*}
  In the limit of a wide charge distribution $\Delta Q^2 \gg \Gamma^2(N_B/N_A)$, this effective Gaussian distribution reduces to $p_A(Q_A)\approx  \frac{N}{N_A}p(Q_A (N/N_A))$. More generally, however, we obtain a Gaussian distribution of enhanced width.

  The computation of the KL-divergence Eq.~\eqref{eq:AverageEntropy} is now reduced to a straightforward Gaussian integral over the distributions $p_A,\Omega_A$, and we obtain  
  \begin{align*}
        S^{\mathrm{a}}&=S_\mathrm{th}-\frac{N_A}{2N}\left(\frac{\Delta Q^2}{\Gamma^2}+\frac{\bar Q^2}{\Gamma^2}-1\right) +\cr 
        &\qquad +\frac{1}{2}\ln\left( 1+\frac{N_A}{N} \left(\frac{\Delta Q^2}{\Gamma^2}-1\right)\right)
  \end{align*}

\subsection{Derivation of the fluctuation entropy} 
\label{sec:derivation_of_the_fluctuation_entropy}

We here compute the dominant fluctuation contributions to the entanglement entropy
for the microcanonical charge distribution from the moments $M^r=\sum \langle
\Psi_{a_1 b_1}\bar \Psi_{a_2 b_1} \Psi_{a_2 b_2}\bar \Psi_{a_3 b_2}\ldots \Psi_{a_r
b_r}\bar \Psi_{a_1 b_r}\rangle$. It will be instructive to first review the
computation for the case considered by Page, where the Gaussian contraction over
random state vectors yields a factor $\langle \Psi_{a b} \bar
\Psi_{a'b'}\rangle=D^{-1}\delta_{a,a'}\delta_{b,b'}$. A contribution of  maximal
number, $r$, of $b$-index summations is obtained by Gaussian pairing $\langle
\Psi_{a_l b_l} \bar \Psi_{a_{l+1}b_l}\rangle=D^{-1}\delta_{a_l,a_{l+1}}$ for all $l$.
This single pairing amounts to taking the average over the distribution prior to
computing the moments, $\langle M^r\rangle \to \langle M\rangle^r$, as discussed in
the first part of the paper. Stepping down in the index order, $N(r,2)=\binom{r}{2}$
terms with $r-1$ summations over $b$ and two over $a$ are obtained by one pairing
outside the above order. The summation over indices for these two terms yields the
estimate $M_r \approx D^{-r}\left(D_B^r D_A + \binom{r}{2} D_B^{r-1}D_A^2\right)$,
and the straightforward differentiation with respect to $r$ gets us to the Page
result, $S\approx N_A\ln 2+\frac{D_A}{2 D_B}$~\cite{Page93}. This observation conveys a number of
messages: first, pairings of generic number of pairing permutations $1<l<(r-2)$ do
not contribute. (Technically, they vanish in the replica
limit $r \rightarrow 1$, up to corrections exponentially small in $D$.) Second, the
single transposition fluctuation contribution, $l=1$, is exponentially small, except
for $N_A\approx N_B$. Finally, the above results becomes wrong for $N_A>N_B$. The
resolution to this problem lies in the inclusion of the opposite limit where just one
$b$-summation results from the likewise unique pairing $\langle \bar
\Psi_{a_{l}b_{l-1}}\Psi_{a_l b_l} \rangle=D^{-1}\delta_{b_{l-1},b_{l}}$. Finally,
symmetry is restored by considering $\binom{r}{2}$ terms with one pairing outside
this scheme, two summations over $b$, $r-1$ over $a$. The differentiation of these
two terms gives $S\approx N_B\ln 2+\frac{D_B}{2 D_A}$, i.e. Page with $A
\leftrightarrow B$.

A closer inspection of the full sum, and of its convergence properties in the limit $r\rightarrow\infty$, shows that the terms with $(r,r-1)$ or $(2,1)$ summations over $b$ need to be kept, depending on whether $D_B$ is larger or smaller than $D_A$. In this way, one obtains the full `Page curve', i.e. the entropy $S\approx N_X\ln 2+\frac{D_X}{2 D_{\bar X}}$ with $(X,\bar X)=(A,B)$ or $(B,A)$ depending on which of the systems is larger.

These general structures remain valid in the case of more general distributions, and specifically the microcanonical one with pairing $\langle \Psi_{a b} \bar \Psi_{a'b'}\rangle=\delta_{Q_a+Q_b,\bar Q}(D\Omega(\bar Q))^{-1}\delta_{a,a'}\delta_{b,b'}$. Substituting these expressions into $M_r$, organizing the sum according to the number of $b$- and $a$-index summations, and trading the $a$ and $b$ summations for summations over $Q_A$ and $Q_B=\bar Q-Q_A$ weighted by spectral densities, we obtain the two alternative representations
\begin{align}
  M_r
&=\frac{1}{F(\bar Q)^r}
\sum_{Q_A,Q_B} \delta_{Q_A+Q_B,\bar Q}\sum_{k=1}^r N(r,k) \times \cr 
& \qquad \times \left\{ \begin{array}{l}
  F_B(Q_B)^{r+1}
\left( \frac{F_A(Q_A)}{F_B(Q_B)}\right)^k \cr 
F_A(Q_A)^{r+1}\left( \frac{F_B(Q_B)}{F_A(Q_A)}\right)^k
\end{array} \right.
\end{align}
with the abbreviations $F(\bar Q)=D\Omega(\bar Q)$, $F_X(Q_X)=D_X \Omega_X(Q_X)$, $X=A,B$, and combinatorial factors $N(r,1)=1$, $N(r,2)=\binom{r}{2}, \ldots$ known as Narayana numbers. Convergence in the limit $r\to \infty$ (more precisely, the analytic properties of the representation of the sum over Narayana numbers in terms of hypergeometric functions~\cite{Liu20,Stanford2019}) implies that the first (second) of these defines an asymptotic $k$-series in the case $F_A(Q_A)<F_B(Q_B)$ ($F_A(Q_A)>F_B(Q_B)$). The terms $k=1,2$ are relevant to the computation of the entanglement entropy, and doing the $r$-derivative we obtain 
\begin{widetext}
\begin{align}
  \label{eq:SWithFlucutations}
  S=-\sum_{Q_A,Q_B} \delta_{Q_A+Q_B,\bar Q} \left[\frac{F_A(Q_A)F_B(Q_B)}{F(\bar Q)}\ln \left(\frac{F_B(Q_B)}{F(\bar Q)}\right)+\frac{1}{2}\frac{F^2_A(Q_A)}{F(\bar Q)}\right]\Theta(F_B(Q_B)-F_A(Q_A))+ (A\leftrightarrow B).
\end{align}
\end{widetext}
If $F_A(Q_A)<F_B(Q_B)$ for all terms in the sum, we are back to the result discussed in the main part of the paper, with corrections in $F_A/F\propto D_A/D_B$ from the second term. However, more sizeable modifications arise for larger subsystems, where swaps $F_A(Q_A)>F_B(Q_B)$ may occur for at least some terms under the sum. 

For the Gaussian spectral densities, Eq.~\eqref{eq:GaussianDoS}, the computation of these sums for general configurations $(\bar Q,N_A)$ is a straightforward if tedious affair. We here limit ourselves to the discussion of equal partitions, $N_A=N_B$, for which the strongest corrections to the previous results will be obtained. In this case, the case distinction in the above sum reduces to 
$\Theta(F_B(Q_B)-F_A(Q_A))=\Theta (|Q_A|-|Q_B|)$, and assuming $\bar Q\ge 0$, this equals $\Theta(Q_A-\bar{Q}/2)$. 

Comparing this expression to $S^\mathrm{a}$ studied in the main text --- the first sum without $\Theta$-function constraint and $F^2_A/F$ correction --- we find the deviation
\begin{align*}
  \delta S \simeq \sum_{Q_A< \bar{Q}/2}\frac{F_A(Q_A)F_A(\bar Q-Q_A)}{F(\bar Q)}\ln \left(\frac{F_A(Q_A)}{F_A(\bar Q-Q_A)}\right),
\end{align*}
where we neglected the non-logarithmic as parametrically subleading (see below). 
Processing the $Q_A$-sum and the spectral density weights as in section~\ref{sec:derivation_of_subsystem_charge_distribution_}, we obtain (see also Ref.~\cite{VidmarRigol17})
\begin{align}
\label{eq:FlucutationCorrection}
   \delta S 
   \simeq -
   \frac{\sqrt{N}}{\sqrt{2\pi }\gamma} \int_{q_A<0} dq_A\,e^{-\frac{Nq^2_A}{2\gamma^2}}
   \left(\frac{N\bar{q}q_A}{\gamma^2}\right)
   =
-\frac{\sqrt{N}\bar{q}}{\sqrt{2\pi}\gamma}.
 \end{align} 
A similar computation shows that the second term in angular brackets in
Eq.~\eqref{eq:SWithFlucutations} yields the correction 
\begin{align}
\delta S
&=
-\frac{1}{2}e^{\frac{N\bar{q}^2}{2\gamma^2}}
{\rm Erfc}\left(\frac{\sqrt{N}\bar{q}^2}{\sqrt{2\gamma^2}}\right)
,
\end{align}
with $\delta S \simeq -\gamma/(\sqrt{2\pi N}\bar q)$ for $\bar{q}\gg \gamma/\sqrt{N}$ and 
cut off  at $\delta S = -1/2$ for $\bar q=0$. We conclude that for generic values $\bar
q \gtrsim \gamma/N^{-1/2}$, Eq.~\eqref{eq:FlucutationCorrection} defines the dominant
fluctuation contribution to the entanglement entropy.

\subsection{A note about distributions} 
\label{sub:what_distributions_actually_}

There are at least three different distributions, 
defined over different support sets, that are relevant for our discussion. 
To define them, 
consider an operator $\hat Q$ with eigenstates $|n\rangle$ and eigenstates $Q_n\in I$ 
inside a certain interval.
\begin{enumerate}
  \item The first (input distribution) is $p(Q)$ some distribution on $I$ one can choose at will.
  \item The second is defined by a density matrix represented in the $\hat Q$ eigenbasis $\hat \rho = \sum_n \rho_n |n\rangle\langle n|$. The set of numbers $\{\rho_n\}$ defines a distribution on the set $\{n\}$ (i.e. according to 
  Born's rule, the probability to measure the outcome $Q_n$ in a measurement of $\hat Q$ is given by $\rho_n$).
  \item The third is a distribution $P(\Psi)$ defined on the Hilbert space of states $\{\Psi\}$, i.e. a complex $D$-dimensional vector space. 
\end{enumerate}
To illustrate how these distributions are related to each other, we first note that distributions 2 and 3 
individually define density matrices as
\begin{align*}
  \hat \rho = \sum_n \rho_n |n\rangle\langle n|,\qquad \hat \rho'= \int D\Psi \,P(\Psi)\,|\Psi\rangle\langle \Psi|,
\end{align*}
where in the second expression $D\Psi = \prod_n d\Psi_n$. 
(Notice that we are here summing over a massively overcomplete set.)
We can then establish a connection between the latter, requiring that the two representations 
$\hat \rho$ and $\hat \rho'$ individually generate the input distribution 1. 
Conceptually, a distribution on the interval, $I$, is generated from one on $\{n\}$ 
by considering $Q_n$ as a random variable and $Q$ as a dependent random variable. 
(Much as e.g. the number parity is a dependent variable of $1,2,3,4,5,6$ when throwing a dice.) 
The distribution of $Q$ is then obtained from that of $Q_n$ as
\begin{align}
     p(Q)=\langle \delta_{Q,Q_n}\rangle_\rho \equiv \sum_n \rho_n \delta_{Q,Q_n}.
   \end{align}   
For example, if $\rho_n = Z^{-1}e^{-\lambda Q_n}$ is a canonical distribution, this becomes 
\begin{align*}
  p(Q)=\frac{1}{Z}\sum_n e^{-\lambda Q_n}\delta_{Q,Q_n}=\frac{D\Omega(Q)}{Z} e^{-\lambda Q},
\end{align*}
with $Z=\sum_n e^{-\lambda Q_n}=D \sum_Q \Omega(Q) e^{-Q}$. In this way $\rho_n$ descends to a distribution $p(Q)$. 
In the same manner, the distribution $P(\Psi)$ generates a distribution $p(Q)$ too,
\begin{align}
  p(Q)&=\langle \delta_{\hat Q,Q}\rangle_\Psi \equiv \int D\Psi P(\Psi) \langle  \Psi |\delta_{\hat Q,Q}|\Psi\rangle. 
\end{align}
Assuming Gaussianity $P(\Psi)=\mathcal N \prod_n e^{-c_n |\Psi_n|^2}$, 
\begin{align}
  p(Q)&
  =\sum_n c^{-1}_n \delta_{Q,Q_n}\equiv D\Omega(Q)c^{-1}(Q),
\end{align}
where in the final step we made the ansatz $c_n = c(Q_n)$. 
If we now impose these induced density distributions to be equal, 
we get the identification
\begin{align}
\label{app_definition_c}
  c(Q)=Z e^{\lambda Q}.
\end{align}
As a sanity check, we notice that the states described by 
the distribution $P(\Psi)$ are normalized (on average),
\begin{align*}
  1&=\langle \langle \Psi|\Psi\rangle\rangle_\Psi 
  \cr 
  &=\int D\Psi P(\Psi)\sum_{n}  |\Psi_n|^2 =\sum_n c_n^{-1}=Z^{-1}\sum_n e^{-\lambda Q_n}.
\end{align*}
Summarizing, there is two a priori different density matrices $\hat \rho$ and $\hat \rho'$ 
descending to the same `macroscopic' charge distribution, and 
what remains to elaborate on their relation. 
To this end, we compare their matrix elements
\begin{align*}
   \hat \rho_{nm}=\langle n|\hat \rho |m\rangle=\frac{1}{Z}\delta_{nm}e^{-\lambda Q_n},
 \end{align*} 
and 
\begin{align}
  \hat \rho'_{nm}&=\langle n|\hat \rho' |m\rangle=\mathcal{N}\int D\Psi e^{-c_n|\Psi_n|^2} \Psi_n\bar \Psi_m
  =\frac{\delta_{nm}}{c_n}.
\end{align}
Employing that 
$c_n=c(Q_n)$ given in Eq.~\eqref{app_definition_c}, 
we conclude that 
\begin{align}
   \hat \rho_{nm}
   &=\hat \rho'_{nm}.
\end{align}
That is, although defined in very different ways, both density matrices are actually equal.

\end{document}